\documentclass[a4paper,10pt]{article}
\usepackage[latin1]{inputenc}
\usepackage{graphicx}
\begin{document}
\begin{titlepage}
\begin{center}
\Large{\bf{Canonical solution of a system of long-range
 interacting rotators on a lattice}}
\end{center}
\vspace{2cm}
\begin{center}
\large{Alessandro Campa$^1$, Andrea Giansanti$^{2, \S}$
 and Daniele Moroni$^2$}
\end{center}
\vspace{1cm}
\begin{center}
\normalsize{$^1$Physics Laboratory, Istituto Superiore di Sanit\`a and
INFN Sezione di Roma1, Gruppo Collegato Sanit\`a\\Viale Regina
Elena 299, 00161 Roma, Italy}
\end{center}
\begin{center}
\normalsize{$^2$Physics Department, Universit\`a di Roma ``La Sapienza'' and
INFM Unit\`a di Roma1, \\ Piazzale Aldo Moro 2, 00185 Roma, Italy}
\end{center}
\vspace{0.5cm}
\begin{center}
\emph{(13 April 2000, accepted for publication in Physical Review E)}
\end{center}
\vspace{1cm}
\begin{center}
\large{\bf {Abstract}}
\end{center}
\vspace{1cm}
\small{
The canonical partition function of a system of rotators (classical
X-Y spins) on a lattice, coupled by terms decaying as the inverse of
their distance to the power $\alpha$, is analytically computed. It is
also shown how to
compute a rescaling function that allows to reduce the model, for any
$d$-dimensional lattice and for any $\alpha<d$, to the mean field
($\alpha=0$) model.
}

\vspace{2cm}
$^{\S}$ \small{Author to whom correspondence should be addressed; electronic address:
Andrea.Giansanti@roma1.infn.it}
\end{titlepage}
\section{Introduction}
Let us consider the following classical hamiltonian model of a system
of rotators:
\begin{equation} \label{num1}
 H=\frac{1}{2} \sum_{i=1}^N L_i^2 \, + \, \frac{1}{2}
\sum_{i,j=1}^N\left[1-\cos(\theta_i-\theta_j)\right]=K+V\, .
\end{equation}
The potential energy $V$ is not thermodynamically stable and the ensemble
averaged energy
density $U=\langle \frac{H}{N}\rangle$ diverges in the thermodynamic
limit (TL) \cite{ruelle}.
If the potential energy term is divided by $N$, then the 
energy density becomes intensive and it 
is bounded as $N$ goes to infinity.

Indeed, dynamics and thermodynamics of the $1/N$ rescaled model
has been extensively investigated \cite{LRR}; in particular, Ruffo and
Antoni, who called it the hamiltonian mean field X-Y model (HMF),
solved it in the canonical ensemble, and compared the
theoretical caloric ($T\, vs\, U$) and magnetization ($M\, vs\, U$)
curves with those obtained from a microcanonical simulation \cite{AR}.

Here we consider a generalization of model (\ref{num1}):
\begin{equation} \label{num2}
 H=\frac{1}{2} \sum_{i=1}^{N} L_i^2 \, + \, \frac{1}{2}
\sum_{i\neq j}^{N} \frac {1-\cos(\theta_i-\theta_j)}
{r_{ij}^\alpha} \, .
\end{equation}
The rotators are placed at the sites of a lattice 
and the interaction between rotators $i$ and $j$ decays as the inverse of
their distance to the power $\alpha$.

A onedimensional version of model (\ref{num2}) has been studied by 
Anteneodo and Tsallis \cite{AT}, who have numerically measured the largest
Lyapounov exponent, as a function of $N$ and $\alpha$.
Through a rescaling factor $N^{*}=\frac {N^{1-\alpha}-1}{1-\alpha}$ 
Anteneodo and Tsallis showed that their results
coincide with those previously obtained for the HMF ($\alpha=0$) model;
this rescaling could then give a well defined TL to model (\ref{num2}). 

In a recent paper Tamarit and Anteneodo, using a rescaling
factor $\tilde{N}= 2^{\alpha}\frac {N^{1-\alpha}-1}{1-\alpha}$, have shown
that the caloric and magnetization curves of model (\ref{num2}) in one
dimension collapse onto the curves 
of the HMF model \cite{TA}.
This universality emerges plotting $T/\tilde{N}$ as a function of
$H/N\tilde{N}$
and $M$ as a function of $H/N\tilde{N}$, from molecular dynamics simulation
of model (\ref{num2}) for different $N$ and $\alpha$ values. These authors
conjecture that 
the results they obtained in the onedimensional case might be general,
valid in any dimension $d$ and for $\alpha <d$, as suggested also
in \cite{TA}.
\section{Partition function}
In this work, inspired by \cite{AR} and \cite{TA}, we analytically compute
the partition function of an $\tilde{N}$-rescaled model (\ref{num2})
for any $d$ and $\alpha <d$.
In formula (\ref{exactn}) we give the right expression of the rescaling
function $\tilde{N}$, to obtain universal state curves for all lattice
models with long range ($\alpha < d$) interactions.

Let us now rewrite the rescaled version of Hamiltonian (\ref{num2}):
\begin{eqnarray}\label{num3}
H&=&\frac{1}{2} \sum_{i=1}^N L_i^2 \, + \, \frac{1}{2 \tilde{N}}
\sum_{i,j=1}^N \frac {1-\cos(\theta_i-\theta_j)}
{r_{ij}^\alpha} \nonumber \\
&&- h_x \sum_{i=1}^N m_{ix} - h_y\sum_{i=1}^N m_{iy} \, ,
\end{eqnarray}
where we have introduced an external magnetic field
$\mathbf{h}=(h_x,h_y)$ of modulus $h$, that makes possible to compute
the magnetization.
The indexes $i,j$ label the sites of a $d$-dimensional generic
lattice; $r_{ij}$ is the distance between them, with periodic
boundary conditions and nearest image convention (the definition of
$r_{ii}$ will be given shortly); $\alpha \geq 0$.
At each site a classical rotator (X-Y spin) of unit
momentum of inertia is represented by conjugate canonical
coordinates $(L_i,\theta_i)$, where the $L_i$'s are angular
momenta, and the $\theta_i$'s $\in [0,2\pi )$ are the angles of
rotation on a family of parallel planes,
each one defined at each lattice point; $x$ and $y$ refer to the
components of boldface twodimensional vectors defined over these planes.
To each lattice site a spin vector
\begin{equation}
\mathbf{m}_i=(m_{ix},m_{iy})=(\cos \theta_i,\sin \theta_i)
\end{equation}
is associated, and the total magnetization is given by:
\begin{equation}
\mathbf{M}=(M_x,M_y)=\frac{1}{N}\sum_{i=1}^N\mathbf{m}_i \, .
\end{equation}
Note in (\ref{num3}) the rescaling factor $\tilde{N}$ in front of the
potential energy term, now written as a free double sum over both indexes.
$\tilde{N}$ should be regarded as an unknown function of
$N,\alpha,d$ and the geometry of the lattice, with the
fundamental property of making
\begin{equation}\label{Ntildedefi}
\frac{1}{\tilde{N}}\sum_{j,j \neq i} \frac{1}{r_{ij}^\alpha}
\end{equation}
an intensive quantity; this guarantees the thermodynamic stability of
the potential. We also note that the sum in (\ref{Ntildedefi}) is
independent of the origin $i$ because of periodic conditions.
To reproduce the usual HMF it is also necessary that
$\tilde{N}(N,\alpha=0,d)=N$.
The constraint $i\neq j$ over the double sum is removed defining
$r_{ii}^\alpha = 1/b$, a finite number. Since the numerator
$1-\cos(\theta_i-\theta_j)$ is zero for $i=j$ the choice of $b$ is
free. The removal of the constraint allows to introduce the distance
matrix $R'_{ij}=\frac{1}{r_{ij}^\alpha}$; the diagonalization of
such matrix is the key point to obtain, in the computation of the
partition function, known integrals in the variables $\theta_i$.

As usual the partition function factorizes in a kinetic
part $Z_K=\left(\frac{2 \pi}{\beta} \right)^{\frac{N}{2}}$, where
$\beta=1/k_B T$, and a potential part $Z_V$. After defining
$R_{ij}=\frac{\beta}{2 \tilde{N}} R'_{ij}$,
$\mathbf{B}=\beta \mathbf{h}$, $C=\exp\left(-\frac{\beta}{2 \tilde{N}}
\sum_{ij} \frac {1}{r_{ij}^\alpha}\right)$, the potential part
can be written as:
\begin{equation}\label{Zmatr}
Z_V=C \int_{-\pi}^{\pi}
 d^N\theta\, \exp\left[\sum_{i,j,\mu}m_{i\mu}R_{ij}
m_{j\mu} + \sum_iB_{\mu}m_{i\mu} \right],
\end{equation}
where $\mu=x,y$. Diagonalizing the symmetric matrix $R=(R_{ij})$ with the
unitary matrix
$U$ such that $R=U^TDU$, $D=(R_i\delta_{ij})$, where $R_i$ are the
eigenvalues of $R$, we can write the first part of the exponent in
(\ref{Zmatr}) as:
\begin{equation}\label{diago}
\sum_{ij}\left(m_{ix}R_{ij}m_{jx} + m_{iy}R_{ij}m_{jy}\right) =
\sum_i\left(n_{ix}^2R_i + n_{iy}^2R_i\right),
\end{equation}
where $n_{i\mu}=\sum_jU_{ij}m_{j\mu}$. In order
to apply the gaussian transformation:
\begin{equation}\label{gauss}
e^{aS^2}=\frac{1}{\sqrt{4\pi a}}\int_{-\infty}^{+\infty}dz e^{-\frac{z^2}{4a}+Sz}
\,\,\,\, a>0
\end{equation}
to each term of the sum in the right hand side of (\ref{diago}), each
$R_i$ must be positive. The spectrum can be explicitly
computed using a $d$-dimensional Fourier transform of matrix $R$, the
eigenvalues being labelled by vectors of the reciprocal lattice. These
eigenvalues are trivially related to those of matrix $R'$. A study of the
spectrum of $R'$ in the limit $N \rightarrow \infty$ and for $b=0$ shows
that: when $\alpha > d$ each element of the spectrum converges to a finite
quantity, the least eigenvalues being negative and of order one in modulus;
when $\alpha < d$ a part of the spectrum converges to a finite quantity,
another part diverges to $+ \infty$, at most as $\tilde{N}$. However this
last part consists of a
fraction of the total number of eigenvalues which goes to zero in the limit
$N \rightarrow \infty$. The least eigenvalue is still negative and of order
one in modulus.
Then part of the spectrum is negative, but it is easily seen that it is
shifted by $b$. Thus calling $p$ the least eigenvalues of $R'$ for $b=0$
and choosing
\begin{equation}\label{defbi}
b=-p + \epsilon \quad \epsilon > 0 \quad,
\end{equation}
we have that with this $b$ the whole spectrum of $R'$ (and therefore that
of $R$) becomes positive. Then for each $i=1,\cdots,N$,
$\mu=x,y$ we can apply (\ref{gauss}) with the correspondence
$a\rightarrow R_i$, $S\rightarrow n_{i\mu}$, $z\rightarrow z_{i\mu}$.
Performing the integrals over variables $\theta_i$ and using the
transformation $z_{i\mu}=2\sum_j(UR)_{ij}\Psi_{j\mu}$ with Jacobian
$2^N\det R$, we can rewrite the partition function as:
\begin{eqnarray}
Z&=&CZ_K\frac{\det R}{\pi^N}\int_{-\infty}^{+\infty} d^N\Psi_x
 d^N\Psi_y  \\
&& e^{ N\left[-\sum_{ij\mu} \Psi_{i\mu} \frac{R_{ij}}
 {N}\Psi_{j\mu} +\frac{1}{N} \sum_l \ln \left( 2 \pi I_0\left( |2 \sum_j R_{lj}
 \mathbf{\Psi}_j+\mathbf{B}|\right) \right) \right] } \nonumber
\end{eqnarray}
where $I_0$ is the zeroth order modified Bessel function.
The isolation of the $N$ factor in the exponential prepares the object
for the use of the saddle point method. The quantity in square brackets
is intensive. Double sums in the first two terms are compensated by
$R/N=(\beta/2 N \tilde{N}) R'$ and the last sum has $1/N$ in front of
it. The argument of $I_0$ is also intensive because involves a
term of the form $\sum_j R_{lj}=(\beta / 2 \tilde{N}) \sum_j R'_{lj}$.
If we call $f(w)$ the function in square brackets, where
$w=(\Psi_{1x},\cdots,\Psi_{Nx},\Psi_{1y},\cdots,\Psi_{Ny})$, then
the application of the method requires the following three conditions:
$f(w)$ admits a stationary point $w_0$; $w_0$ is a simple stationary
point, i.e., $\det He f|_0\neq 0$, where $He f|_0$ is the hessian matrix
of $f$ in $w_0$; the path of integration can be
deformed (generally going into ${\cal C}^{2N}$) into a path that
passes through $w_0$ following the steepest descent of $f(w)$ and such
that $f(w)<f(w_0)$ throughout the all path. If the point $w_0$ is a
maximum no deformation is necessary and the method is also called the
Laplace method. Since, as we show below, $w_0$ is indeed a real-valued
maximum, we readily obtain for the free energy per particle $F$:
\begin{eqnarray}\label{fren}
-\beta F&=&\lim_{N\rightarrow \infty}\frac{\ln Z}{N}=
\lim_{N\rightarrow \infty} \{ \frac{1}{2}\ln
\left( \frac{2\pi}{\beta}\right) -\frac{\beta}{2\tilde{N}}\sum_{j}
\frac{1}{r_{ij}^\alpha} \nonumber \\
&&+ \max_{w}[f(w)] +\frac{1}{N}\ln \frac{\det R}{\sqrt
{\det \left( -\frac{N}{2} He f|_0\right)}} \} \, .
\end{eqnarray}
The stationary point $w_0$ is given by the vector
$(\Psi_x,\cdots,\Psi_x,\Psi_y,\cdots,\Psi_y)$, homogeneous on the lattice
sites. Defining $\mathbf{\Psi}=
(\Psi_x,\Psi_y)$, its direction is that of $\mathbf{B}$, and its modulus
$\Psi$ is given by the solution of:
\begin{equation}\label{psieq}
\Psi =\frac{I_1}{I_0}\left( \beta\left[A\Psi +h\right] \right) \, ,
\end{equation}
with
\begin{equation}\label{ascal}
A=\frac{1}{\tilde{N}} \sum_j R'_{ij}=\frac{1}{\tilde{N}} \left[ b +
 \sum_{j\neq i}\frac{1}{r_{ij}^\alpha} \right] \, ,
\end{equation}
and where $I_1$ is the first order modified Bessel function. In (\ref{ascal})
$A$ does not depend on $i$ because of the periodic boundary conditions.
We note that when $h=0$ we have
infinitely many degenerate solutions, since only the modulus $\Psi$ is
determined. Evaluation of the elements of the hessian matrix at the
stationary point gives:
\begin{equation}\label{hesmat}
\left. -\frac{N}{2} \frac{\partial^2 f}{\partial \Psi_{i\mu}
\partial \Psi_{j\nu}}
\right|_0 = 
R_{ij} \delta_{\mu \nu} - (R^2)_{ij} g_{\mu \nu}(w_0)
\end{equation}
where we do not give the explicit expression of $g_{\mu \nu}(w_0)$.
As we will see shortly, the eigenvalues analysis of the hessian matrix
(\ref{hesmat}) shows that the stationary point $w_0$ is a maximum. Then,
Laplace method applies and Eq. (\ref{fren}) is valid. However, only in the
long range case ($\alpha<d$) the last term in the rightmost side of
(\ref{fren}) is zero; when $\alpha>d$ its expression does not appear to
be manageable. We will comment on this point later. Restricting then to
$\alpha<d$, and computing the derivative of (\ref{fren}) with
respect to the magnetic field we find that the magnetization
$M=\langle|\mathbf{M}|\rangle$ is given by the solution $\Psi$ of
(\ref{psieq}). Then the internal energy $U$ is given by:
\begin{equation}\label{energy}
U=\frac{\partial (\beta F)}{\partial \beta} = \frac{1}{2\beta} +
\frac{A}{2}(1-M^2) -hM \, .
\end{equation}
Equations (\ref{psieq}) and (\ref{energy}) are the same as those
of HMF, as soon as a proper $\tilde{N}$ rescaling gives
\begin{equation}\label{rescal}
A=\frac{1}{\tilde{N}} \sum_j R'_{ij}=\frac{1}{\tilde{N}} \left[
 b + \sum_{j\neq i}\frac{1}{r_{ij}^\alpha} \right] = 1 \, .
\end{equation}
Now, from equations (\ref{hesmat}) and (\ref{rescal}), and calling
$\lambda_n$ the eigenvalues of $R'$, we find that, choosing $\mathbf{B}$
along one of the coordinate axes, the eigenvalues of the hessian matrix
at the stationary point are given by:
\begin{eqnarray}\label{eighes}
\chi^{(1)}_n&=&\frac{\beta}{2}\frac{\lambda_n}{\tilde{N}}\left[1-
\left(\beta-\Psi^2\beta-\frac{\Psi}{\Psi+h}\right)
\frac{\lambda_n}{\tilde{N}}\right] \\
\chi^{(2)}_n&=&\frac{\beta}{2}\frac{\lambda_n}{\tilde{N}}\left[1-
\frac{\Psi}{\Psi+h}\frac{\lambda_n}{\tilde{N}}\right] \quad \quad
n=1,\cdots,N
\nonumber
\end{eqnarray}
Following our previous analysis we have that:
\begin{equation}
\frac{\epsilon}{\tilde{N}}\leq \frac{\lambda_n}{\tilde{N}}\leq 1 \, .
\end{equation}
Then we immediately see that $\chi^{(2)}_n$ are all positive for any
$\beta$ and $h$; for $\chi^{(1)}_n$ we need to include $\Psi (\beta,h)$ from
(\ref{psieq}). We have checked numerically that the quantity in round brackets
in (\ref{eighes}) is always smaller than $1$, and therefore $\chi^{(1)}_n$
are also all positive. From (\ref{eighes}) we can derive an expression for
the determinant of matrix (\ref{hesmat}). It is given by:
\begin{eqnarray}\label{dethes}
\frac{1}{N}&&\ln \det \left( -\frac{N}{2} He f|_0\right)=\frac{2}{N}\ln \det R
\nonumber \\ +&&\frac{1}{N}\sum_{n=1}^N \{ \ln \left[1-
\left(\beta-\Psi^2\beta-\frac{\Psi}{\Psi+h}\right)
\frac{\lambda_n}{\tilde{N}}\right]\nonumber \\ +&&\ln \left[1-
\frac{\Psi}{\Psi+h} \frac{\lambda_n}{\tilde{N}}\right]
\} \, .
\end{eqnarray}
When $\alpha<d$ most of $\frac{\lambda_n}{\tilde{N}}$ go to zero for
$N\rightarrow \infty$, then the sum in (\ref{dethes}) is
effectively constituted by the terms with the remaining
$\frac{\lambda_n}{\tilde{N}}$. These terms are a fraction of $N$ that, as
we already pointed out, goes to zero when $N\rightarrow \infty$. If we call
$N'(N)$ this fraction, then the sum in (\ref{dethes}) can be bounded
from above by $\frac{N'}{N}c \rightarrow 0$ for $N\rightarrow \infty$, where
$c$ is a finite number. Therefore the last term in (\ref{fren}) is zero.
When $\alpha>d$ all terms contribute to the sum in (\ref{dethes}), and we
can not give a meaningful expression for (\ref{fren}). At the end of the
calculations we can let $\epsilon \rightarrow 0$ in (\ref{defbi}).

Then we have shown that any model with $\alpha<d$ on any lattice is
equivalent to HMF. From (\ref{rescal}) we get an exact expression
for $\tilde{N}$:
\begin{equation}\label{exactn}
\tilde{N} = -p + \sum_{j\neq i}\frac{1}{r_{ij}^\alpha} \, .
\end{equation}
We have made a microcanonical simulation of Hamiltonian
(\ref{num3}) on a threedimensional simple cubic lattice
in zero magnetic field, using a fourth
order simplectic algorithm \cite{yosh} with time step $0.02$, selected to have
relative energy fluctuations not exceeding $1/10^6$. We have chosen a fixed
$N=343=7^3$, and have simulated various energy densities
$H/N$ and various $\alpha<3$. In Fig. 1 we show that the numerical
caloric curves collapse onto the universal HMF curve. The kind of results
shown in \cite{TA} for a onedimensional lattice, where a slightly
different $\tilde{N}$ has been used.

\begin{figure}[htbp]
\begin{center}
\includegraphics[width=12cm]{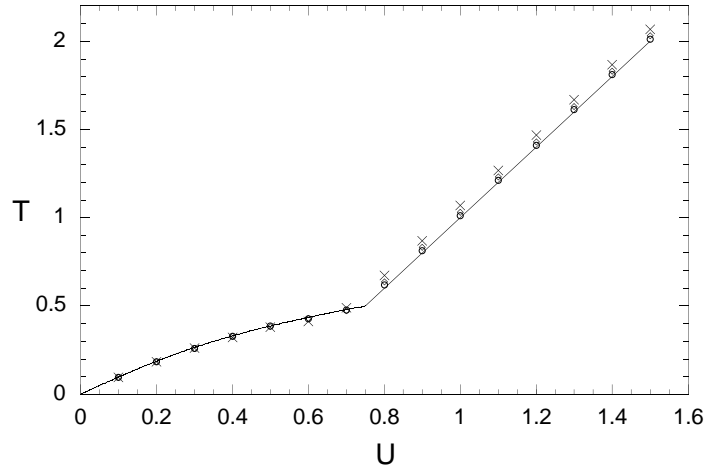}
\end{center}
\caption{The full line gives the canonical theoretical caloric curve
(temperature $T$ $vs$ energy density $U$) for
long range rotators compared with the microcanonical simulation of a
threedimensional simple cubic lattice for three different $\alpha$ values:
$0.75$ (open circles), $1.5$ (diamonds) and $2.25$ (crosses). Note that
in spite of the size of the system, still not very large (side with $7$
lattice sites), the results already follow very well the theoretical
curve.}
\label{Fig. 1}
\end{figure}

\section{Conclusions}
Going back to the beginning of our discussion:
it is now clear that model (\ref{num2}) completely reduces to
model (\ref{num1}) for $\alpha=0$. In model (\ref{num1}) the range of the
interactions is infinite; each rotator interacts with
all the others and with the same intensity. To get
a well defined TL it is sufficient to divide
$V$ in (\ref{num1}) by $N$, the total numbers of rotators.
It is then possible to compute caloric and magnetization
curves \cite{AR}; the spatial  arrangement of the rotators has no effect on
them since the intensity of the interaction is the same for each couple
of rotators. In this work
we have shown that, when considering model (\ref{num2}), it is possible
to take into account the spatial $d$-dimensional arrangement
of the rotators and the decaying of their mutual interaction
through a factor $\tilde{N}$, which is computable for any
periodic lattice and any $\alpha < d$. Dividing by $\tilde{N}$
the potential energy in (\ref{num2}), the model gets a well defined
TL and it is possible to compute state curves which become those
of the HMF model with a proper normalization of the constant $A$
in (\ref{rescal}). The HMF ($\alpha =0$) model has revealed peculiar 
equilibrium and nonequilibrium properties \cite{LRR}, namely:
ensemble inequivalence, metastability, collective oscillations,
anomalous diffusion and interesting chaotic properties, both in
the ferromagnetic and antiferromagnetic case. On the basis of
the thermodynamical equivalence here established it would be
interesting to investigate the $\alpha$ dependence of all these
properties. The study of the Lyapounov exponents in \cite{AT} 
is the first in this direction.
\section{Aknowledgments}
A. G. warmly thanks C. Tsallis for having suggested the study
of the long range interacting rotators.


\begin{thebibliography}{9}
\bibitem{ruelle}
D. Ruelle, {\it Statistical mechanics: rigorous results},
(Addison-Wesley, New York 1989).
\bibitem{LRR}
V. Latora, A. Rapisarda, and S. Ruffo, cond-mat/0001010,
to appear in Prog. Theor. Phys. Supplement.
\bibitem{AR}
M. Antoni and S. Ruffo, Phys. Rev. E {\bf 52}, 2361 (1995).
\bibitem{AT}
C. Anteneodo and C. Tsallis, Phys. Rev. Lett. {\bf 80}, 5313 (1998).
\bibitem{TA}
F. Tamarit and C. Anteneodo, Phys. Rev. Lett. {\bf 84}, 208 (2000).
\bibitem{yosh}
H. Yoshida, Phys. Lett. A {\bf 150}, 262 (1990).
\end{thebibliography}
\end{document}